\documentclass[%
 aip,
 amsmath,amssymb,
 reprint,%
]{revtex4-1}

\usepackage{graphicx}%
\usepackage{dcolumn}%
\usepackage{bm}%

\usepackage[utf8]{inputenc}
\usepackage[T1]{fontenc}
\usepackage{mathptmx}
\usepackage{etoolbox}
\usepackage{todonotes}
\usepackage{hyperref}
\usepackage{appendix}
\makeatletter
\def\@email#1#2{%
 \endgroup
 \patchcmd{\titleblock@produce}
  {\frontmatter@RRAPformat}
  {\frontmatter@RRAPformat{\produce@RRAP{*#1\href{mailto:#2}{#2}}}\frontmatter@RRAPformat}
  {}{}
}%
\makeatother

\newcommand{\ffig}[1]{Figure~\ref{#1}}

\newcommand{\erf}[1]{\ensuremath{\operatorname{erf({#1})}}}

\begin{document}

\preprint{AIP/123-QED}

\title[]{Excitable response of a noisy adaptive network of spiking lasers}
\author{S. Barland}
\affiliation{Universit\'e C\^ote d'Azur-CNRS, Institut de Physique de Nice, France}

\author{O. D'Huys}%
\affiliation{Department of Advanced Computing Sciences, Maastricht University, Netherlands}

\author{R. Veltz}
\affiliation{Inria Center at Universit\'e C\^ote d'Azur, Cronos team, France
}%

\date{\today}%

\begin{abstract}
We analyze experimentally and theoretically the response of a network of spiking nodes to external perturbations. The experimental system consists of an array of semiconductor lasers that are adaptively coupled through an optoelectronic feedback signal. This coupling signal can be tuned from one to all to globally coupled and makes the network collectively excitable. We relate the excitable response of the network to the existence of a separatrix in phase space and analyze the effect of noise close to this separatrix. We find numerically that larger networks are more robust to uncorrelated noise sources in the nodes than small networks, in contrast to the experimental observations. We remove this discrepancy considering the impact of a global noise term in the adaptive coupling signal and discuss our observations in relation to the network structure.
\end{abstract}

\maketitle

\IfFileExists{./modtime.tex}{\input{modtime.tex}}{}

\section{lead paragraph}

Complex real world systems are often modeled as networks of interacting nodes. The collective dynamics of such networks is strongly determined by the dynamics of the individual nodes but also by their couplings and the different forms of disorder which may exist within the network. The case of spiking nodes is particularly relevant due to the prominent role of neural excitability and collective phenomena in high or low level cognition tasks. Beyond the case of static networks, whose topology and coupling strengths are constant in time, \textit{adaptive} networks, whose coupling strength evolves with the state of the network itself, are currently attracting attention, in spite of the challenges presented by their analysis. Finally, far from the idealized case of a perfectly homogeneous network operating under perfectly controlled conditions, most real networks include both quenched disorder and various sources of noise. Here, we study experimentally, numerically, and theoretically the excitable response of an adaptive network of semiconductor lasers. All lasers are coupled via a slowly evolving optoelectronic feedback, such that they can be effectively coupled or not depending on the network state and history. We discuss how excitability can be an emerging property of the network for some specific coupling, and the existence of a separatrix that governs the excitable dynamics at the network level. Finally, we analyze how network heterogeneity and different sources of noise are necessary to provide a complete theoretical picture of the experimental observations: uncorrelated noise in the nodes pushes them out of unstable fixed points and global noise in the coupling variable brings correlations across the network.

\section{\label{sec:intro}Introduction}

The dynamics of complex networks is at the center of many high-impact research areas including climate science, complex systems engineering, and neuroscience, to name a few (see, e.g. \cite{yanchuk2021dynamical} for a recent overview on the topic). Due to its prominence in neuroscience, one area of particular interest is the collective dynamics of spiking elements, with emphasis on stochastic effects \cite{zaks2005noise,orr2021noise} or topology \cite{khaledi2018variability} or both \cite{sonnenschein2013excitable,masoliver2017coherence}. In addition, in the context of physics-based neuromorphic computing \cite{marcucci2020theory,markovic2020physics}, spike-based machine intelligence is currently emerging as an expressive and energy-efficient paradigm building on a variety of hardware platforms based on electronic but also on photonic implementations\cite{roy2019towards,skalli2022photonic,romeira2023brain}. 

Several purely photonic spiking architectures have been analyzed in recent years, including lasers and microresonators \cite{dolcemascolo2018resonator,kelleher2021optical,tiana2023quantifying,jha2022photonic,skalli2022photonic}.  However, it is often hard to couple many optical or optoelectronic nodes. Although theoretical work presents promising approaches \cite{perego2016collective,lamperti2017disorder,alfaro2020pulse}, experimental architectures rarely go beyond delayed self-coupling \cite{garbin2015topological,romeira2016regenerative,terrien2017asymmetric,robertson2019toward} or very few nodes \cite{yacomotti2002coupled,kelleher2010excitation,deng2017controlled,feldmann2019all,robertson2019toward}. At the interface between photonics and electronics, optoelectronic systems present remarkable advantages due to their flexibility \cite{al2010excitability,inagaki2021collective} and their capacity to operate back and forth conversions between optical and electronic signals \cite{hejda2022resonant}.

In the following, we focus on the effects of coupling on the excitable dynamics of a noisy and heterogeneous network of semiconductor lasers with nonlinear optoelectronic feedback \cite{dolcemascolo2020effective}. One of the key features of this network is that the coupling between lasers takes place through a (slowly) evolving variable. Thus, contrary to most networks where nodes are dynamic but coupling strengths are static, this system can be interpreted as an adaptive network, which may bring very peculiar phenomena such as Canard cascades \cite{balzer2024canard} and Canard resonance \cite{d2021canard}. Such adaptive networks \cite{berner2023adaptive}, while often used in the modeling of natural \cite{gross2006epidemic,meisel2009adaptive,sawicki2022modeling} or man-made systems \cite{berner2021adaptive,ciszak2021collective,ciszak2024type}, are comparatively much less experimentally studied.

Here, we study the response to external perturbation of homogeneous or heterogeneous networks with different kinds of adaptive couplings in the presence of noise. Without feedback, the semiconductor lasers are ON (lasing) or OFF (not lasing), depending on the parameters. The network becomes excitable due to the adaptive coupling. In the case of sum-coupling, we show that this effect appears as the network size increases, thus showing a form of emergent excitability \cite{ciszak2020emergent}.  
When comparing a population of single nodes with feedback that are excitable (a setup reminiscent of a single layer in a feedforward neural network) to a globally coupled network, we observe that excitability is lost on the ensemble in presence of noise, but that mean-intensity coupling can restore collective excitability.

Finally, we analyze the impact of different noise sources on the statistical properties of the excitable response of a heterogeneous network to perturbations. In section \ref{sec:exp} we present the experimental arrangement and in section \ref{sec:obs} the experimental observations. In section \ref{sec:model}, we present a mathematical model of the experiment and analyze the existence of a separatrix in phase space, both for the single node with feedback and at the network level. In \ref{sec:simuls}, we perform numerical simulations which shed light on the role of noise sources in the experimental observations. We discuss our results in section \ref{sec: discussion} and provide a summary in section \ref{sec: summary}.

\section{Experimental system and connection control}
\label{sec:exp}

\begin{figure}[h]
	\centering
	\includegraphics[width=0.5\textwidth]{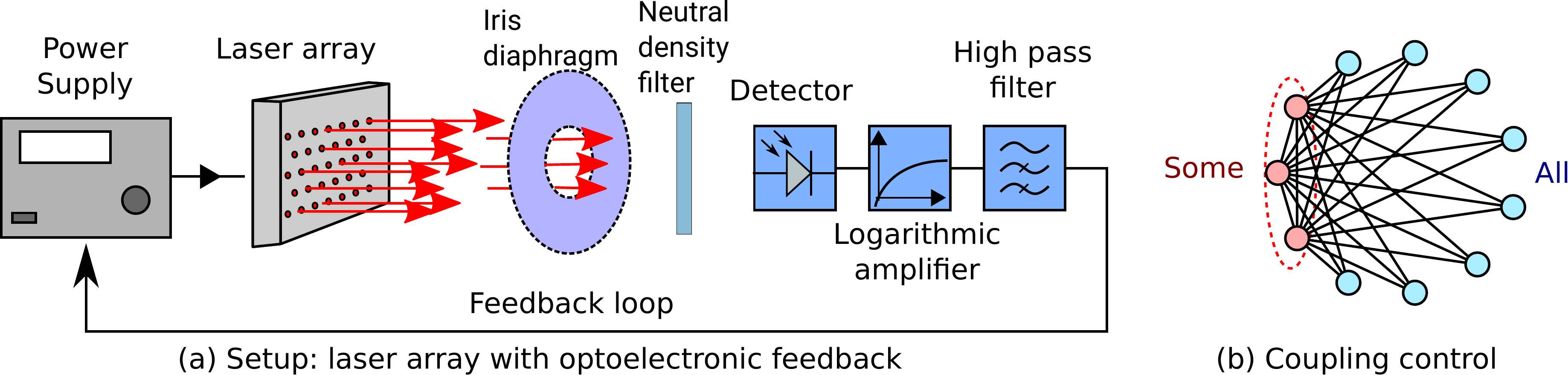}
	\caption{Operating principle of the experiment (a) and connectivity control (b). By opening or closing the iris diaphragm, a subpopulation can be selected for coupling to the optoelectronic feedback loop. The neutral density filter allows for sum or mean intensity coupling.
	\label{fig:setup}}
\end{figure}

The principle of the experimental arrangement (inspired by \cite{al2010excitability}) is shown in \ffig{fig:setup}. It is based on an array of semiconductor lasers, adaptively coupled through a nonlinear optoelectronic feedback. We briefly describe the key points of the experiment here for self-consistency and refer the reader to \cite{dolcemascolo2020effective,d2021canard,dolcemascolo2018semiconductor} for a more detailed description. An array of 451 Vertical Cavity Surface Emitting Lasers emits mutually incoherent light beams. These are steered towards a photodetector which converts the optical intensity into a voltage signal. This signal is logarithmically amplified to provide a saturable nonlinearity and high-pass filtered so as to remove its continuous component. The resulting voltage is fed back into the control input of the power supply driving the whole array. On the optical path between the laser array and the photodetector, an iris diaphragm can be fully opened (in which case the sum of all laser intensities reaches the photodetector) or closed to different degrees. In the latter case, a sub-population (down to one single individual laser) illuminates the photodetector and therefore enters the optoelectronic feedback loop. Between the iris and the detector, a neutral density filter can be used to scale the amount of light reaching the photodetector. When the iris is closed around a unique beam, this single laser drives the whole population. Opening the iris around three lasers leads to this sub-population of three driving the array via the \textit{incoherent sum} of their light intensities. Tuning the neutral density filter (or equivalently the transamplifier gain of the photodetector) to downscale this sum by a factor three leads to the \textit{mean intensity} of this sub-population driving the array. In the following, we refer to these two situations respectively as \textit{sum-} and \textit{mean-intensity} coupling. 

The system is a slow-fast system with the optoelectronic feedback being the slowest variable. We refer the reader to \cite{dolcemascolo2020effective,d2021canard} for a more complete description of the different dynamical regimes but the key observation is that increasing the drive current of the laser array beyond their coherent emission threshold leads the system towards a relaxation oscillation state in which it collectively and periodically oscillates between the branches of its critical manifold. Near this oscillatory state, we have described the excitable dynamics of a single node in \cite{d2021canard}. Here, we focus on excitability at the network level.

\section{Experimental observations}
\label{sec:obs}

\subsection{Sum coupling: From non-excitable nodes to an excitable network}
\label{subsec:size}

Due to the simplicity of the experimental setup, we can effectively analyze how the number of nodes coupled back into the network impacts its overall dynamics. In the following, we experimentally analyze how the system’s response to external perturbations varies with the number of nodes coupled back into the network, while keeping all other parameters constant. We apply a drive current just below the the coherent emission threshold, so that the non-lasing state is stable, (and no spikes are triggered by noise). We perturb the network \textit{globally} by shining a short pulse of light of 20 microseconds duration with controllable intensity on the photodetector in the feedback loop. We monitor the voltage signal which corresponds to the sum intensities of the perturbation pulse and of the selected sub-population. We perform this experiment for seven values of the optical power of the perturbation and for each value we record the results over 1000 independent realizations to assess the statistical significance of the observations.

\begin{figure}[h]
	\centering
	\includegraphics[width=0.5\textwidth]{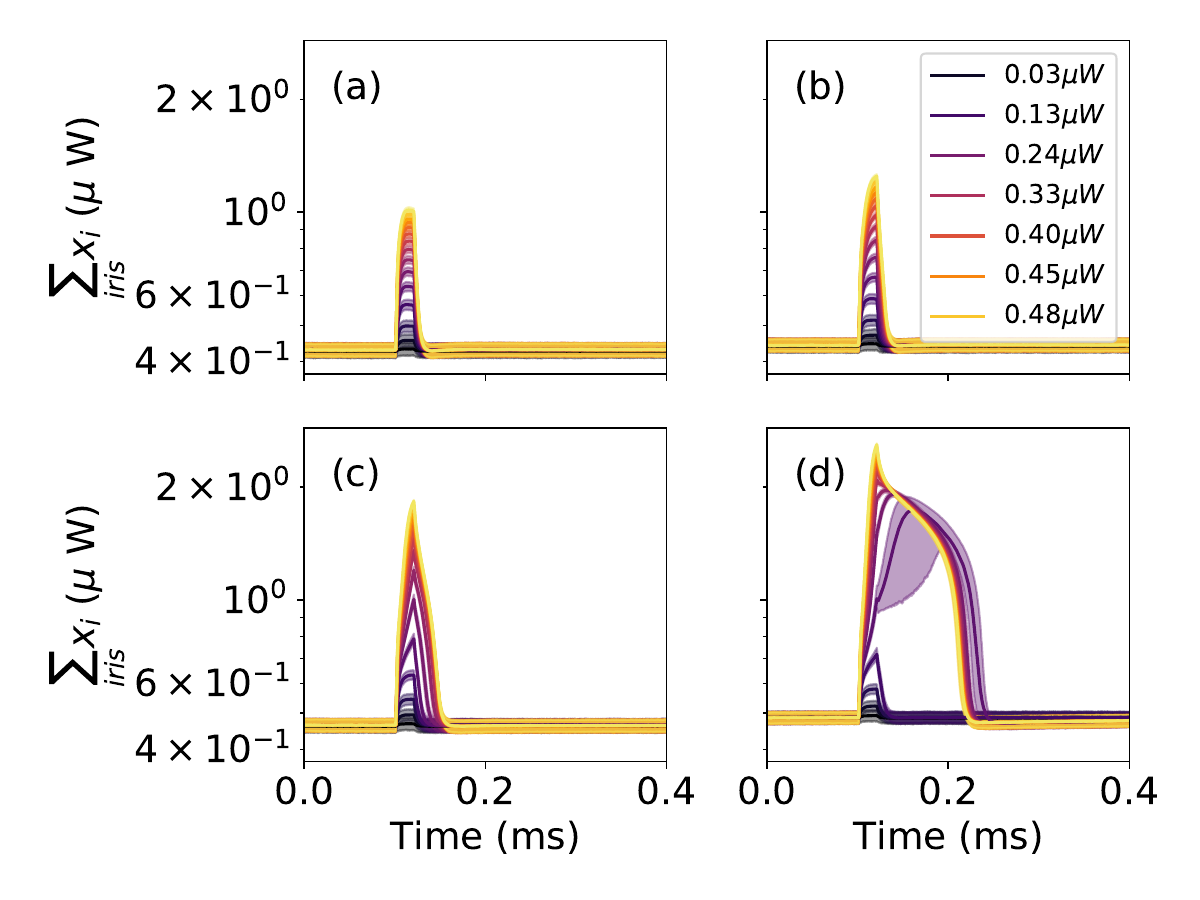}
	\caption{Response to perturbations depending on the number of nodes in the feedback loop:  (\textbf{a}) 7, (\textbf{b}) 19, (\textbf{c}) 37, (\textbf{d}) 60 nodes. When the number of nodes is sufficiently large (\textbf{d}), the network as a whole becomes excitable.
	\label{fig:size}}
\end{figure}   

The results are shown on \ffig{fig:size}. For each power, the curves show the average over the 1000 realizations, surrounded by a shaded area corresponding to the 5-95\% probability boundaries (\textit{i.e.} only 10\% of the trajectories fall outside of that area).

When a small number of nodes is coupled back (panel (a), seven nodes) the measured signal corresponds essentially to a linear response of the system to the perturbation followed by a relaxation towards the stable state. For a larger number of nodes (panel (b) and even more so panel (c)) the response of the system becomes increasingly nonlinear as is apparent from its shape, which differs noticeably from the one in (a). In all these cases, there is essentially no dispersion of the trajectories around their mean value, the residual width being mostly attributed to detection noise.  Finally, for the largest number of nodes (panel (d), 60 nodes) the clear signature of excitability appears: perturbations smaller than 0.24$\mu W$ elicit a mostly linear response while above this threshold large excursions are observed. Close to that threshold, the trajectory distribution is broader pointing at the sensitivity of the system to noise in this regime. We underline that, for small network size (cases (a,b,c)), even applying much stronger perturbations does not elicit an excitable response. We have checked that by applying perturbations up to 2~$\mu$W, an optical power equivalent to the maximal optical power emitted by the network in case (d), at the peak of the excitable response.

From the above, one concludes that when the number of nodes is sufficiently large, the network collectively behaves as an excitable system. This shows that in this case, excitability is an emerging property of the network. As we discuss below in section \ref{sec:simuls} however, this emergence is not observed in the mean coupling configuration and it can be immediately interpreted as a result of the \textit{sum-} coupling.

\subsection{Ensemble average of uncoupled nodes vs mean-intensity coupled network}

We now compare an ensemble of nodes, each with adaptive self-feedback, and an adaptive mean-intensity coupled network. The nodes each are excitable due to the feedback, but they are not coupled to each other - this can be thought of as a single layer of a feedforward neural network. We compare this layer to a mean intensity coupled network, based on the same type of nodes and coupling parameters. We apply an external perturbation, and measure the response in presence of temporal noise. 

\begin{figure}[h!]
	\includegraphics[width=0.5\textwidth]{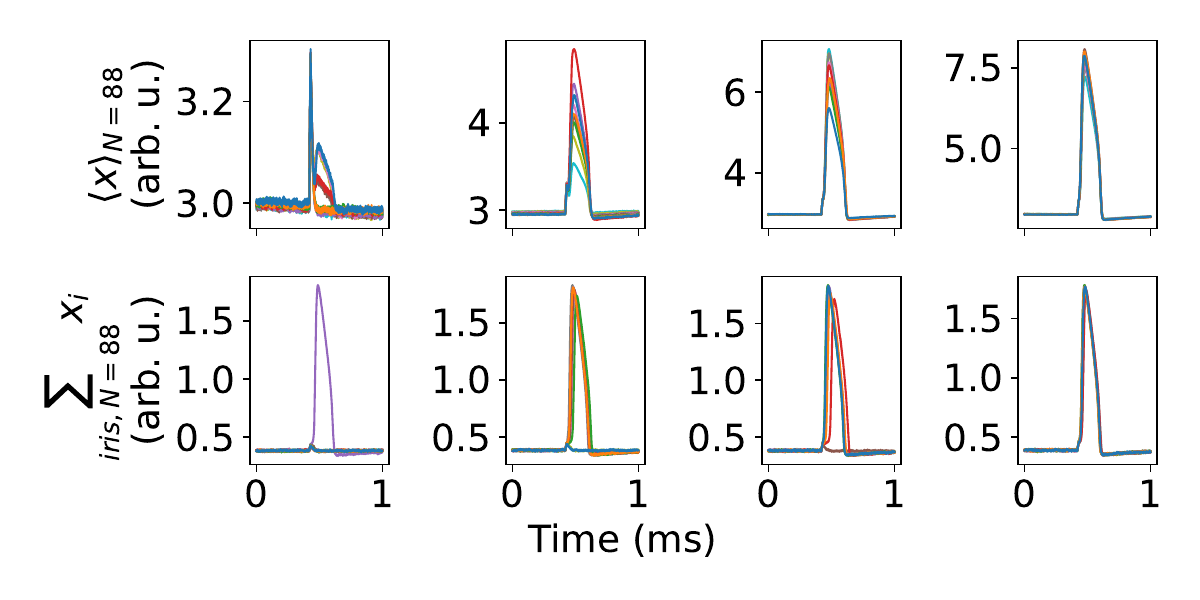}
	\caption{Ensemble average response of independent nodes with adaptive feedback vs response of fully connected adaptive network to external perturbation in the excitable regime. Top row: ensemble average of 88 nodes with feedback in response to increasing external perturbations, 10 ensemble averages are shown in each case. Bottom row: response of a fully connected network of 88 coupled nodes to increasing external perturbations, 10 realizations are shown in each case.
	\label{fig:convergence}}
\end{figure}

To measure the response of a single laser with feedback, we close the iris so that only one laser is coupled back to the driver controlling the whole array, as was done in \cite{d2021canard}. From there, the mean response of an ensemble of size $N$ can be obtained by averaging over $N$ independent realizations of an experiment. In a completely deterministic system, this ensemble, which consists of perfectly identical nodes, would respond identically, leading to the emission of an excitable spike if and only if the perturbation is sufficient to overcome the excitability threshold. Instead, in presence of noise, the average response of the ensemble differs markedly from the response of a single node. On the top row of \ffig{fig:convergence}, we plot for four increasing perturbations strengths ten responses of a ensemble with $N=88$. On the first panel (perturbation power 0.24$\mu W$), one distinguishes an initial fast pulse (which corresponds basically to the linear response of a single node with feedback when no excitable orbit is triggered), sometimes followed by a longer and variable return to the original rest state. On the second panel (0.25$\mu W$), the linear part of the response is still visible but it is followed by a larger, yet highly variable excursion. This larger excursion then becomes dominant and less variable for the largest two perturbations (third and fourth panel, 0.27 and 0.31$\mu W$). The origin of this behavior is clearly related to the presence of noise, which makes the excitability threshold crossing for each node a stochastic process. In this case, for intermediate perturbation strength, the response of the ensemble is affected by absence of response of some nodes and (to a lesser extent) by variations in the response time of the nodes. Thus, at the ensemble level, the excitability property of the nodes is lost due to the presence of noise and the ensemble does not collectively behave as an excitable system.

On the contrary, a mean-intensity coupled network consisting of $N = 88$ nodes does show collective excitability, as shown on the bottom row of \ffig{fig:convergence}. For this experiment, the iris which selects the part of the population which couples back to the whole network is open so that the light emitted by 88 lasers reaches the feedback detector. In this case, the dynamics of these 88 lasers rule that of the whole array and we observe the dynamics of the 88 selected lasers. As previously, we show for four increasing perturbation strengths ten examples of the network responses. Contrary to the previous case, only two types of responses are observed: either linear relaxation to the original stable state or a large excursion with very low variability between realizations. Due to the presence of noise (also at the network level), some responses are skipped and some are delayed but, contrary to the case of the ensemble, the network is collectively excitable due to the very fast convergence towards the slow manifold of the network \cite{dolcemascolo2020effective,d2021canard}.

\subsection{Response to perturbations: one node with feedback vs network}

\begin{figure}[h!]
	\includegraphics[width=0.5\textwidth]{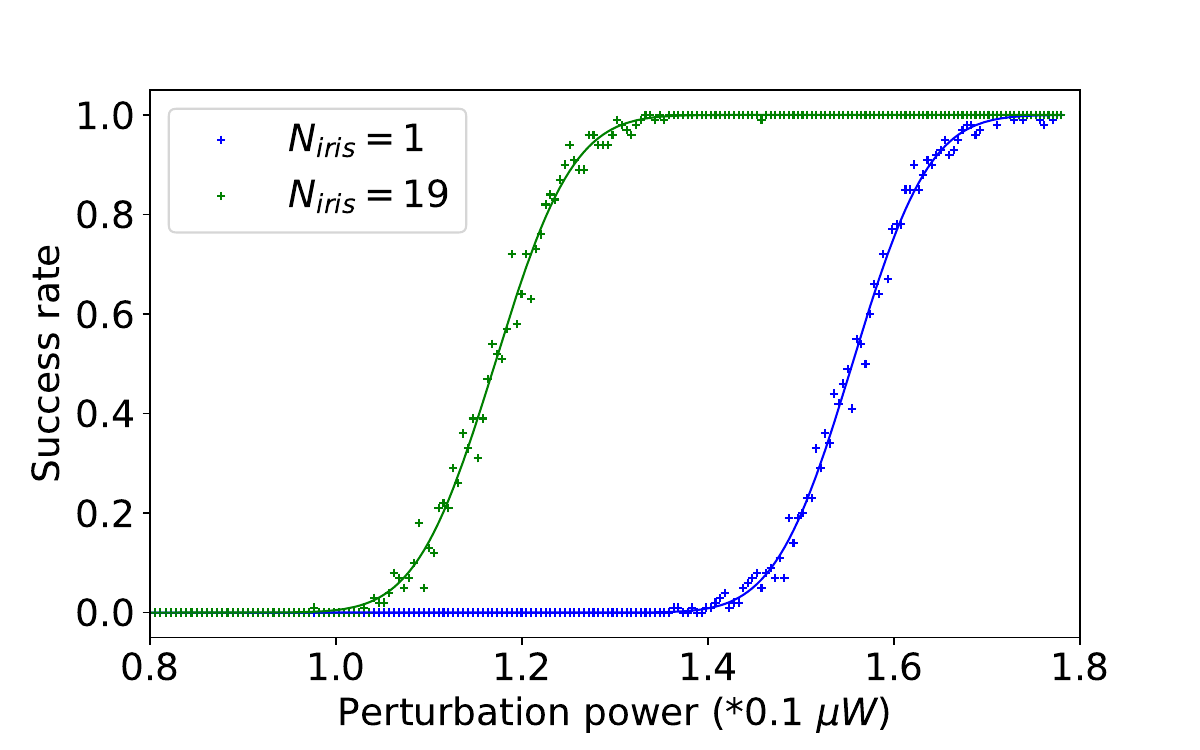}
	\caption{Experimental measurement of the efficiency of perturbations in triggering an excitable response for one single laser with adaptive feedback and when 19 lasers adaptively coupled via their mean intensity. In both cases the efficiency is measured over 100 independent realizations.
	\label{fig:respcurve}}
\end{figure}

It is clear from the previous observations that the dynamics of the system (node with adaptive feedback or adaptively coupled network) in response to perturbations is not fully deterministic. We analyze this effect in more detail by measuring statistically the efficiency of perturbations as is usually done experimentally to characterize excitability. This response curve is shown on \ffig{fig:respcurve} for a single node and for a mean-intensity coupled network. To obtain these curves, we first close the iris around one single laser and tune the bias current and neutral density filter close to but below the slow-fast relaxation oscillation regime, where excitability is expected. We then apply 100 perturbations and count the number of spikes emitted in response to these perturbations. The same procedure is repeated over 200 values of the perturbation power, from 0.08 to 0.18~$\mu W$. For the network case, we open the iris around 19 lasers and tune the neutral density filter to divide the total intensity across the iris by 19 to reach the mean-intensity coupling. Then, we tune the current value so as to place the network in the excitable regime and we repeat the whole procedure of applying series of perturbations of different powers. In all cases, care has been taken to separate the perturbations sufficiently in time so that they can be considered as independent realizations of the experiment. The curves for $N_{iris}=1$ and for $N_{iris}=19$ both show unequivocally the existence of an excitability threshold, situated at slightly different values of the perturbation power. Both curves clearly indicate that the emission of a spike is not a purely deterministic process and the effect of noise can be observed from the steepness of the transition from 0 to 1 in the efficiency. 

In order to quantify the stochastic part of the dynamics, we fit the sigmoid function:
\begin{equation}
S = \frac{1}{2}[1-\erf{\lambda(P-P_0)}]
\label{eq:sigmoid}
\end{equation}
to both response curves, where $S$ it the perturbation success rate and $P_0,\lambda$ are the excitability threshold and scale parameter. This equation has only been derived formally for the response of a single excitable element close to a saddle-node on a circle bifurcation \cite{pedaci2011} and we do not attempt to establish it in our much more complex (high dimensional) case. Nevertheless, it fits very well all of our data and as such it is useful to quantify via the parameter $\lambda$ how the probability of triggering a spike depends on $P$ and comparing this dependence in different configurations.

We find that $\lambda_{N=1}=10.8\pm1*10^{-2}$ and $\lambda_{N=19}=10.6\pm0.3$. Thus, up to the precision of the fits, the network of 19 nodes and a single node with feedback have identical sensitivity to noise.

\begin{figure}[h]
	\includegraphics[width=0.5\textwidth]{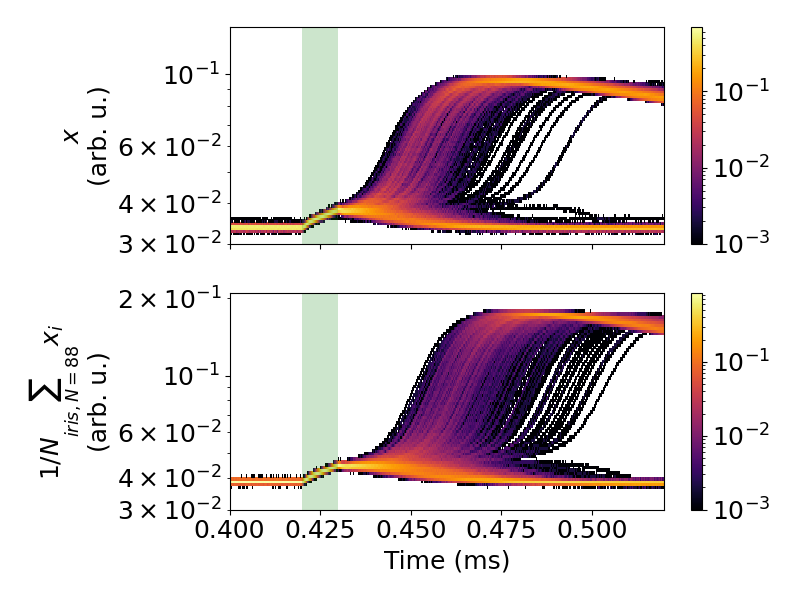}
	\caption{Experimental observation of near threshold distribution of the trajectories of a single node with feedback (top) and a mean intensity coupled network (bottom). Both distributions are very similar but, perhaps unexpectedly, the dispersion of the network trajectory is slightly larger than that of a single node. The $k$ and $\Delta_0$ parameters have been chosen such that the perturbation has approximately 0.5 efficiency.
	\label{fig:departure}}
\end{figure}

We further analyze the transition region between 0 and 1 efficiency by observing the different realizations for the perturbation power such that the efficiency is close to 0.5. We compare on \ffig{fig:departure} two configurations, $N_{iris}=1$ (top) and $N_{iris}=88$ (bottom). At time 0.42~ms, the optical pulse serving as perturbation is switched on and at time 0.43~ms it is switched off. From that time on, both the single node with feedback and the adaptive network follow unsteered trajectories driven by their phase space structure and intrinsic noise. In both cases, one observes relaxation to the original state or departure towards the excitable orbit. In both cases however, a few trajectories remain for a very long time (up to 0.04~ms) in a central region where the final orbit seems to be still undetermined. These trajectories suggest that either systems remain very close to a separatrix where the dynamics is slow and ultimately noise driven. Most striking is the fact that there are here no obvious differences in the behavior of a unique node with feedback and that of a large network. On the one hand, it suggests the existence of similar separatrices in both cases. On the other hand, it raises questions about the role of different noise sources and diversity in this system, as one could expect that independent noise sources in the case of the network would lead to a much weaker sensitivity to noise. Instead, we see that the distributions of the trajectories for a single node with feedback and for a network are essentially identical. As we shall see in the following sections, this observation can be fully explained by the structure of the separatrix of the network and the role of distinct noise sources.

\section{Phase space analysis}
\subsection{Model}
\label{sec:model}
In the following, we briefly remind the model and analyze the deterministic origin of excitability first at the single node level and second at the network level, showing that in both cases a separatrix exists. We then perform numerical simulations which highlight the role of the distinct noise sources in the triggering of an excitable pulse in response to perturbations.

The physical model for a single laser with feedback described by its emitted intensity $s$, population inversion $n$ and the optoelectronic induced current modulation $I$ that constitutes the adaptive feedback, reads\cite{dolcemascolo2020effective}:

\begin{eqnarray}
\dot{s} & = & \big[g(n-n_t)-\gamma_0\big]s \\\
\dot{n} & = & I_0+kI - \gamma_cn - g(n-n_t)s \\\
\dot{I} & = & -\gamma_fI+\dot{f_f(s)} 
\end{eqnarray}
where $g, n_t, \gamma_0$ describe the stimulated emission gain factor, the carrier transparency value and the intensity relaxation rate respectively. $I$ represents the pumping current and $\gamma_c$ the population decay rate. The optoelectronic signal is characterized by a linear gain term $k$, a slow relaxation rate $\gamma_f$ and a saturable (logarithmic) function of the emitted intensity $f_f(s)$.

The dimensionless model equations for $N$ lasers indexed by $i$ after proper scaling ($x=\frac{g}{\gamma_c}s$, $y=\frac {g}{\gamma_0}(n-n_t)$ and $w=I-f_f(\frac{\gamma_c}{g}x)$) are:

\begin{equation}
\label{eq:laserN}
\begin{aligned} 
\dot{x_i} &=x_i\left(y_i-1\right) \\ 
\dot{y_i} &=\gamma\left(\delta_i-y_i+k(w+Af(\langle x \rangle+P(t)))-x_i y_i\right) \\ 
\dot{w} &=-\epsilon(w+Af(\langle x \rangle+P(t)))\,,
\end{aligned}
\end{equation}
where $\langle x\rangle =\frac1N\sum\limits_{i=1}^Nx_i$. The external perturbation pulse is modeled as $P(t) = P\cdot\mathbf 1_{[t_0,t_1]}(t)$ and, like in the experimental setup, it is added to the light in the optoelectronic loop. We choose the same saturable nonlinearity as in the experiment $f(x) = \ln\left( 1+\alpha x\right)$. Here, $(x(t),y(t))$ represent the dynamics of the individual lasers, while the variable $\dot{w}(t)$, that is common for the whole network, represents the slowly varying adaptive coupling strength, driven by the mean intensity $\langle x \rangle$.

We model spontaneous emission noise by supplementing Eq.\eqref{eq:laserN} with an additive noise term in the evolution of the complex laser electric field $E_i$, $\lvert E_i(t)\rvert^2 = x_i(t)$.

\begin{equation}
dE_i = \frac12\left(1+i\alpha_H\right)E_i(y_i-1)+\sigma dB_i\,,
\end{equation}
where $B_i$ are independent Brownian processes and $\sigma$ is the noise strength. We include the linewidth enhancement factor $\alpha_H$ for completeness, but it does not play a role in observed dynamics.

In addition to spontaneous emission noise, we also model electronic noise in the feedback detector and subsequent amplifiers with an additional stochastic term. The third equation in \eqref{eq:laserN} becomes
\begin{equation}
    dw = -\epsilon(w+Af(\langle x \rangle+P(t)))dt + \sqrt\epsilon\sigma_w dB_w\,,
\end{equation}
where $B_w$ is a Wiener process and $\sigma_w$ quantifies the electronic noise strength (scaled by $\sqrt{\epsilon}$ for time scale consistency).

\subsection{Single-node spiking dynamics}
In this section, we provide a geometric understanding of the spiking dynamics of a laser with feedback, \textit{e.g.} $N=1$. The phase space structure of this model has already been analyzed in \cite{dolcemascolo2020effective}, and we summarize the main results here, which is required for the analysis of the excitable dynamics. Not taking into account the perturbation, for $\delta<1$ the system has a stable non-lasing OFF equilibrium at $X_0^{(1)}=(0,\delta,0)$. Hence, at $t=t_0$, the system is close to this equilibrium. 

Since $\epsilon$ is small, we consider the limit $\epsilon\to 0$ and study the system in the slow-fast framework \cite{fenichel_geometric_1979}. The slow manifold consists of two branches: the OFF branch is given by $x=0, y(w)=\delta +kw$. The OFF branch is attracting for $w<\frac{1-\delta}{k}$. On this branch, $w(t)$ and $y(t)$ move towards the stable non-lasing fixed point $(0,\delta,0)$. 

On the ON branch, we have $y=1$. The intensity $x(w)$ solves the equation \begin{equation}
x=\delta-1+k(w+Af(x)) \label{eq:equil-1}\,,
\end{equation}
which has two roots $x^{(1)}_{sl}<x_1^{(1)}$ for $w$ large enough; the latter root forms the attracting part of the ON branch, the former is the repelling part. For $\delta<1$, there is no fixed point on the ON branch. 

The spiking follows typical excitable dynamics: triggered by the pulse, the system may reach the attracting ON branch. On the ON branch, $w(t)$ and $x(t)$ slowly decrease, until the minimal value of $w$ is reached. At this point, the system jumps back to the OFF branch, and returns slowly along the OFF branch to the fixed point.

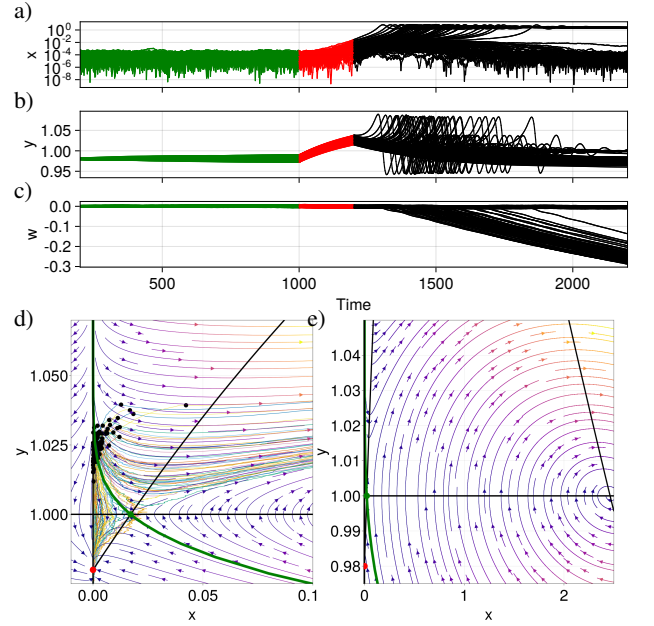
\begin{figure}[h]
	\setlength{\unitlength}{1cm}
	\begin{picture}(8, 8)
		\put(0,4){\includegraphics[width=0.47\textwidth]{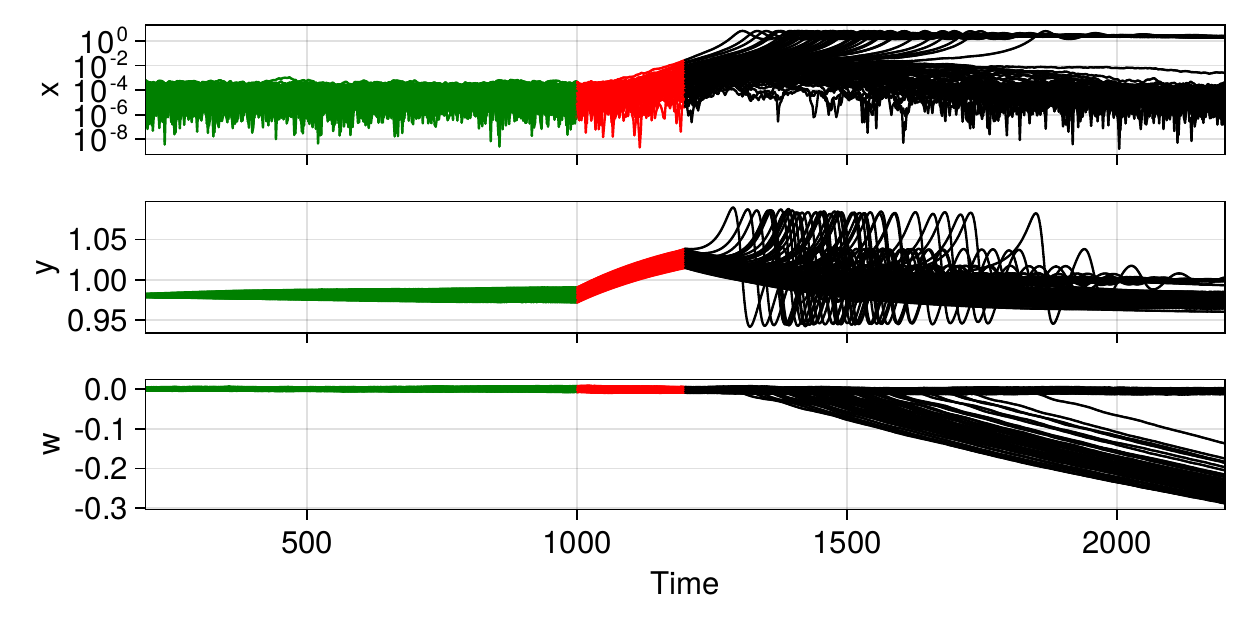}}
		\put(4,0){\includegraphics[width=0.23\textwidth]{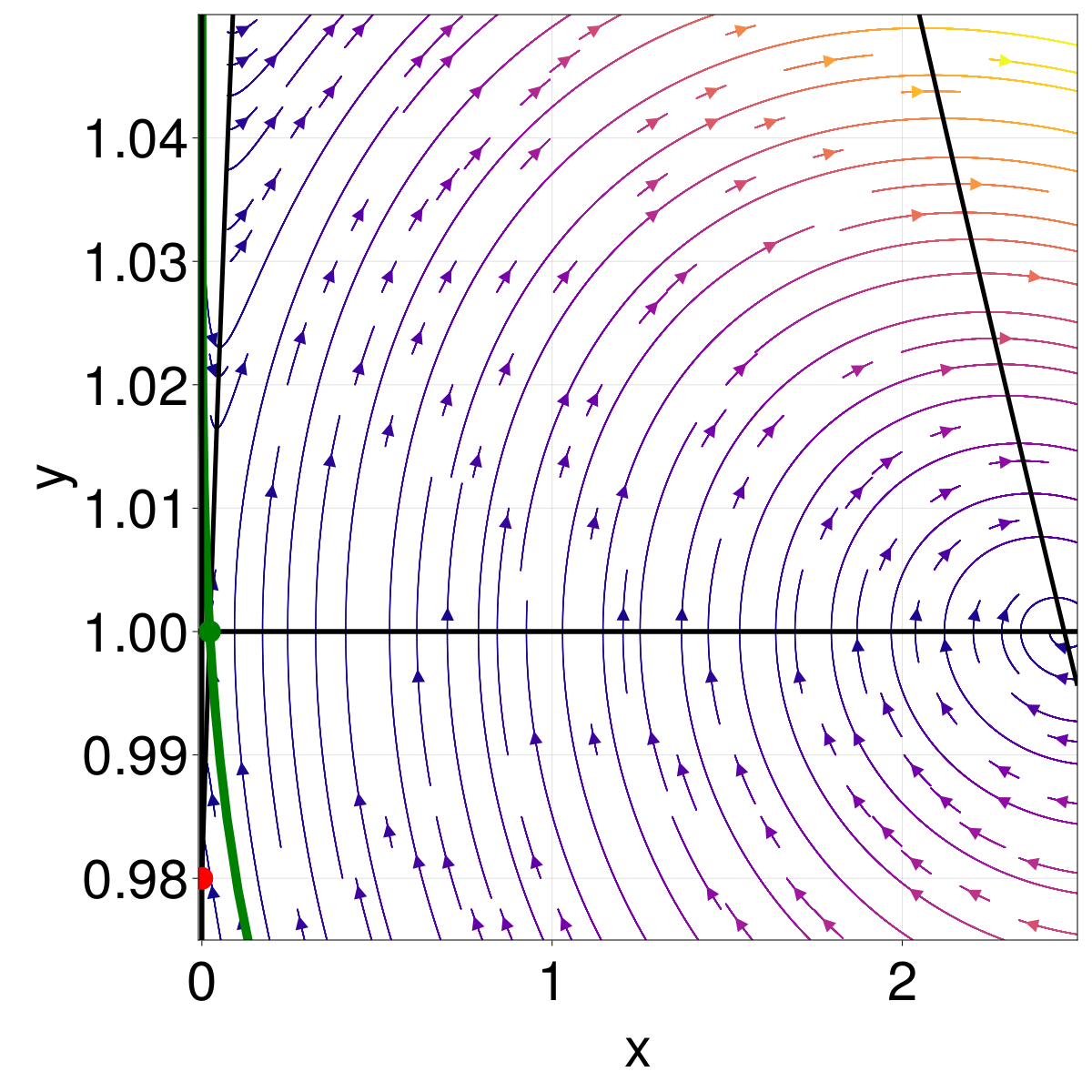}}
		\put(0,0){\includegraphics[width=0.23\textwidth]{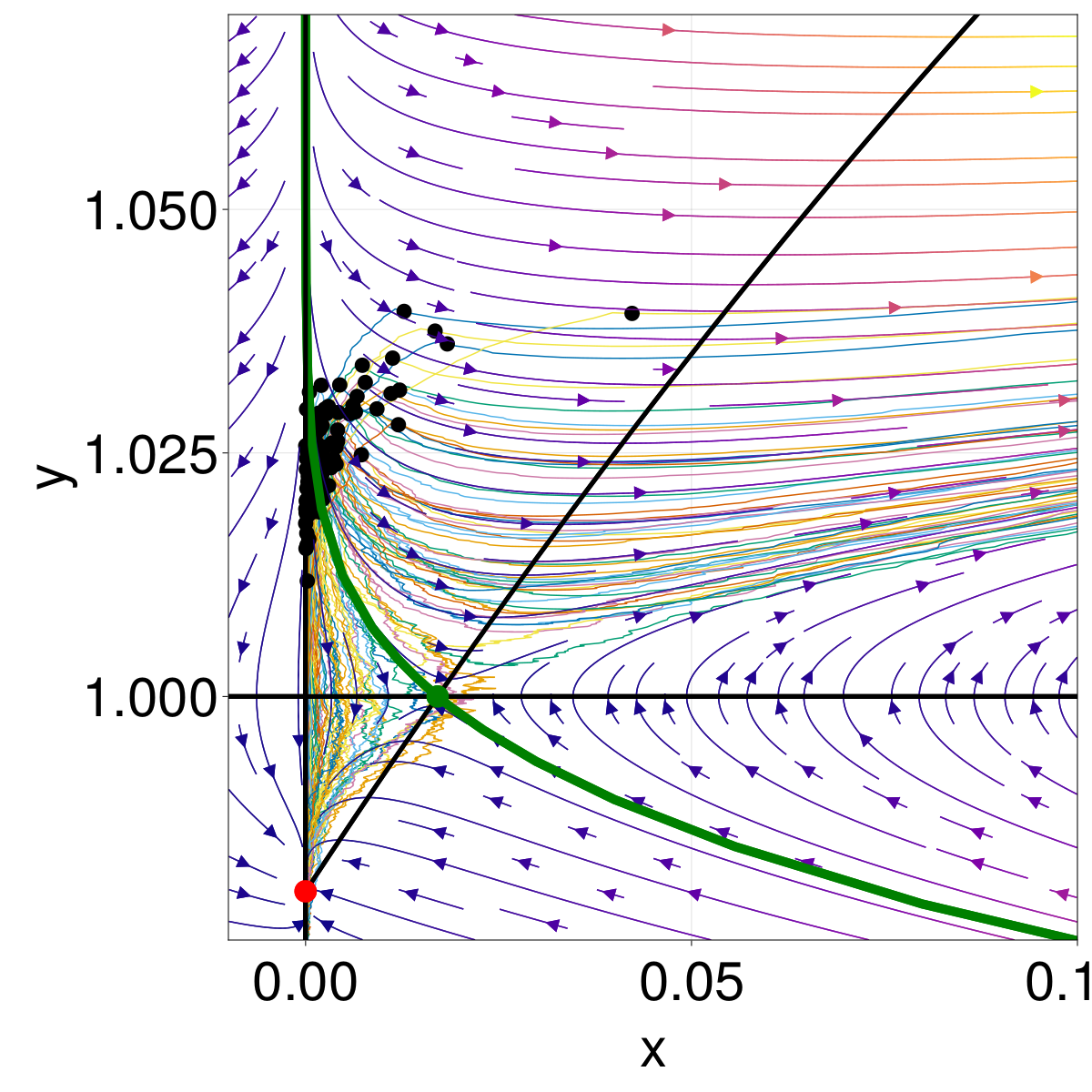}}
		\put(0.1,8.1){a)}
		\put(0.1,6.9){b)}
		\put(0.1,5.7){c)}
		\put(0.1,4.0){d)}
		\put(4,4.0){e)}
	\end{picture}
	\caption{Top: 100 realizations $t_1-t_0 = 200 a.u.$. Left/Right: Vector field for $\epsilon = w=0$. Nullclines are in black. The red dot is $X^{(1)}_0$. The equilibrium near $x\approx 2.5$ is $X^{(1)}_1$. The green dot is the saddle fixed point $X^{(1)}_{sl}$ with its stable manifold displayed in green. The parameters are $\alpha_H = 4, k = 1.7, \delta = 0.98, \gamma = 0.004, A = 1.43, \sigma = 0.001, P = 0.039, \alpha_f = 0.9, N = 1, \epsilon = 0.0002, \sigma_w=0.007$. 100 realizations. }
	\label{fig:vf}
\end{figure}

Whether the system spikes in response to the pulse depends on the noise in the light intensity as we discuss below. We show timetraces of the system in response to a perturbation pulse in Fig. \ref{fig:vf}(a-c). The green part, before the pulse, shows that the system is close to the stable non-lasing fixed point. Due to spontaneous emission noise, we see that $x>0$. Indeed, $x=0$ is an invariant manifold of the system, and without accounting for spontaneous emission noise, the perturbation (in the shape that we apply it) cannot cause an excursion $x>0$. During the pulse, indicated by the red timetraces, we can observe that $y(t)$ increases towards $y=\delta+kAf(P)$ (but it does not reach this new equilibrium value), while $x(t)$ remains small, but starts an average increase as $y>1$. The pulse is short compared to the slow time scale of the system ($\epsilon(t_1-t_0)\ll 1$), and indeed $w(t)$ does not change during and immediately after the pulse.

Therefore, we analyze the spiking dynamics in the fast flow for $w=0$. Beside the OFF equilibrium $X_0^{(0)}=(0,\delta+kw)$, the fast flow has two more equilibria for which the laser is ON (e.g. $x>0$ and $y=1$) corresponding to branches of the slow manifold. Thus, we have the following equilibria where $X^{(1)}_0,X^{(1)}_1$ are stable and $X^{(1)}_{sl}$ is a saddle point:
\begin{align*}
X^{(1)}_0 &=(0, \delta+kw),\\
X^{(1)}_{sl} &=(x^{(1)}_{sl}, 1),\\
X^{(1)}_1 &=(x_1^{(1)},1).
\end{align*}
The stable manifold of $X^{(1)}_{sl}$ serves as a separatrix between the basins of attraction of the two stable equilibria. We show the vector field, along with this separatrix, in Fig. \ref{fig:vf}(d,e). The black dots show the position of the system in phase space at $t=t_1$, for 100 realizations. They clearly show that that the pulse brings the system from the neighborhood of the stable equilibrium (indicated by the red dot) to the vicinity of the separatrix. The noise then makes the dynamics escape, or not escape the basin of attraction of the stable equilibrium $X^{eq}_0$, a paradigm intensively studied by Freidlin and Wentzel \cite{freidlin_random_2012} in the small noise limit $\sigma\to0$.

\subsection{Spiking at the network level}

Let us have a look at the general case with $N$ lasers (Eq. \eqref{eq:laserN}). We refer to earlier work \cite{d2021canard} for a more complete analysis of the branches of the slow manifold. If all $\delta_i<1$, the OFF equilibrium $x_i=0, y_i=\delta_i, w=0$ is stable. Without a perturbation, the system remains in the vicinity of this fixed point.

Also here, we interpret the spiking dynamics in terms of the fast flow $\epsilon \to 0$. For the fast flow, the system has a multitude of equilibria, as each laser can be ON or OFF. However, for identical lasers, only two equilibria are stable: the OFF equilibrium, and an equilibrium where all lasers are ON; we refer to the appendix for proof. We conjecture that the $2N-1$-dimensional hyperplane that separates their basins of attractions is the stable manifold of the symmetric saddle point between them.

If all lasers are ON ($y_i=1$), we find the equation
\begin{equation}
\label{eq:equil-N}
x_i=\delta_i-1+k(w+Af(\langle x \rangle))\,,
\end{equation}
which gives
\[
\langle x \rangle=\langle \delta \rangle-1+k(w+Af(\langle x \rangle))\,,
\]
and this is very similar to the equation we studied in the case $N=1$ (Eq. \eqref{eq:equil-1}). We thus find two values for the mean intensity, satisfying
\[
\langle x \rangle = \frac1N\sum\limits_{i=1}^N x_i = x^{(N)}_{sl}\ (\mbox{resp.}\quad x^{(N)}_{1})
\]
where $x^{(N)}_{sl},x^{(N)}_1$ are defined analogous to  the previous section. Using \eqref{eq:equil-N}, this gives two symmetric equilibria $X^{(N)}_{sl},X^{(N)}_{1}$ for which all lasers are ON. The separate intensities $x_i$ are then given by $x_i = \langle x \rangle +\delta_i-\langle \delta \rangle$.

For a mean-intensity coupled network of identical lasers, the stability of all fixed points is calculated in the appendix - a similar picture holds if the lasers are sufficiently similar (\textit{i.e.} different $\delta_i$  values) \cite{kato2013perturbation}. Just like for a single laser, $X^{(N)}_1$ is a stable equilibrium for the fast flow. We find that $X^{(N)}_{sl}$ is a saddle point, with one unstable direction and $2N-1$ stable directions. We conjecture that the stable manifold of this saddle point acts as a separatrix.

\begin{figure}[h]
	\setlength{\unitlength}{1cm}
	\begin{picture}(8, 3.5)
		\put(0,0){\includegraphics[width=0.21\textwidth]{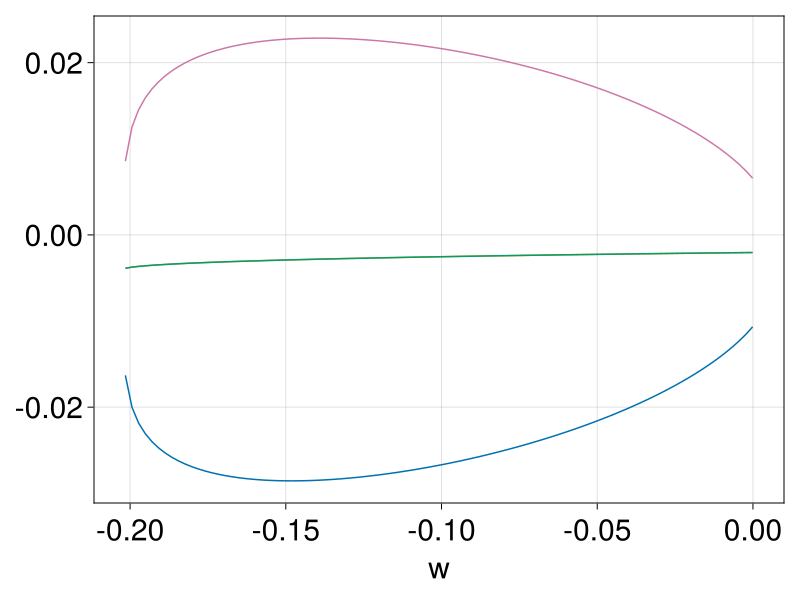}}
		\put(4,0){\includegraphics[width=0.21\textwidth]{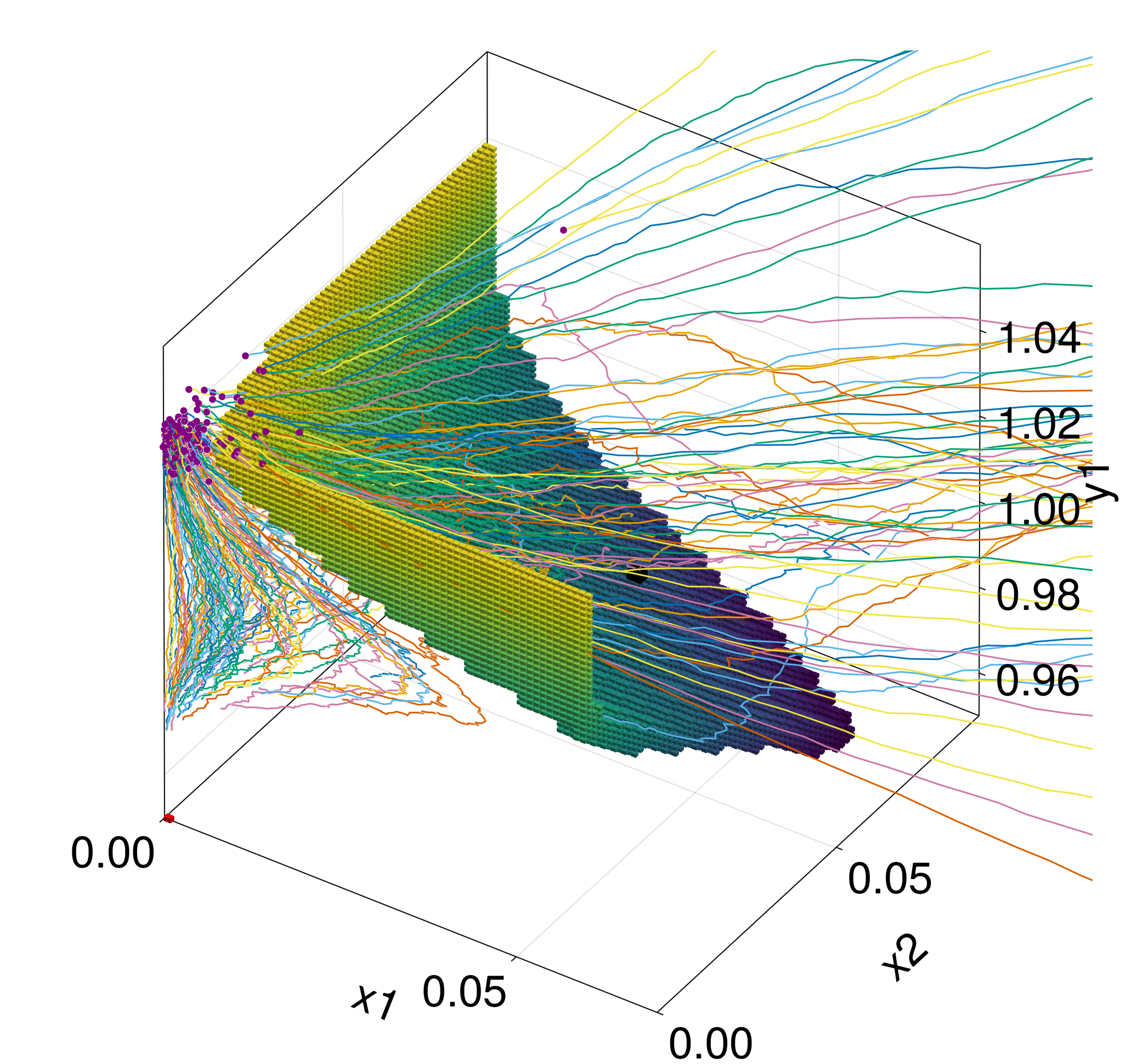}}
		\put(0,3.2){a)}
		\put(4,3.2){b)}
	\end{picture}
	\caption{Left: real part of the eigenvalues (green is double) of the jacobian at the equilibrium $X^{(2)}_{sl}$ as function of $w$. Right: Separatrix for $\epsilon = w=0$ and $N=2$ projected on $\{y_2=0\}$. The red dot is $X_0^{(2)}$. The black dot is the projection of $X_{sl}^{(2)}$ on $\{y_2=0\}$. The parameters are $\alpha_H = 4.0, k = 1.7, \delta = [0.97, 0.99], \gamma = 0.004, A = 1.25, \sigma = 0.001, P = 0, \alpha_f = 0.9, P = 0.078$. 100 realizations, visualized with \cite{DanischKrumbiegel2021}.}
	\label{fig:vfN=2}
\end{figure}

We demonstrate the dynamics numerically for the case $N=2$.
The equilibria $X^{(2)}_{sl}$ can be computed using BifurcationKit \cite{bk} for different values of $w$ and have a one dimensional unstable manifold (see Figure~\ref{fig:vfN=2}~Left). The nonlinear stable submanifold of $X^{(2)}_{sl}$ is 3d locally being an hyperplane because the number of eigenvalues with negative real part is $3$. At this point, we propose that this manifold serves as a separatrix between the numerically observed stable equilibria $X^{(2)}_{0}$ and $X^{(2)}_{1}$ as in the case $N=1$. 

Although computing the stable manifold is numerically challenging, the above hypothesis can be confirmed by using the following alternative approach. We compute the separatrix using a bisection method by sampling the trajectories which converge to $X^{(2)}_{1}$ in Figure~\ref{fig:vfN=2}~Right. Doing so, we observe that, up to numerical accuracy, $X^{(2)}_{sl}$ belongs to this separatrix. We thus conjecture that the separatrix and the stable manifold are the same, at least close to $X^{(2)}_{0}$.

From the above, we conclude that the situation in the case $N=2$ is very similar to the case $N=1$ in the sense that also when two lasers are coupled, there exists a true separatrix between the linear relaxation and the excitable orbit for the dynamics of the whole network in the $2N+1$ dimensional space. We conjecture that this holds true for general $N$.

\section{Numerical simulations}

\label{sec:simuls}

Starting from the geometrical and theoretical analysis reported above, we attempt to reproduce numerically the experimental observations of the response of the mean-intensity coupled network to external perturbations. We first focus on the role of the properties of the network itself (size and diversity among the nodes) and second on the distinct sources of noise. 

As we shall see, the response curves become steeper with N if the Brownian processes $B_i$ are uncorrelated, in contradiction with the experimental observations. This contradiction can not be explained by diversity in the network but it can be solved by taking into account a physically plausible global noise source.

\begin{figure}[h]
	\includegraphics[width=0.5\textwidth]{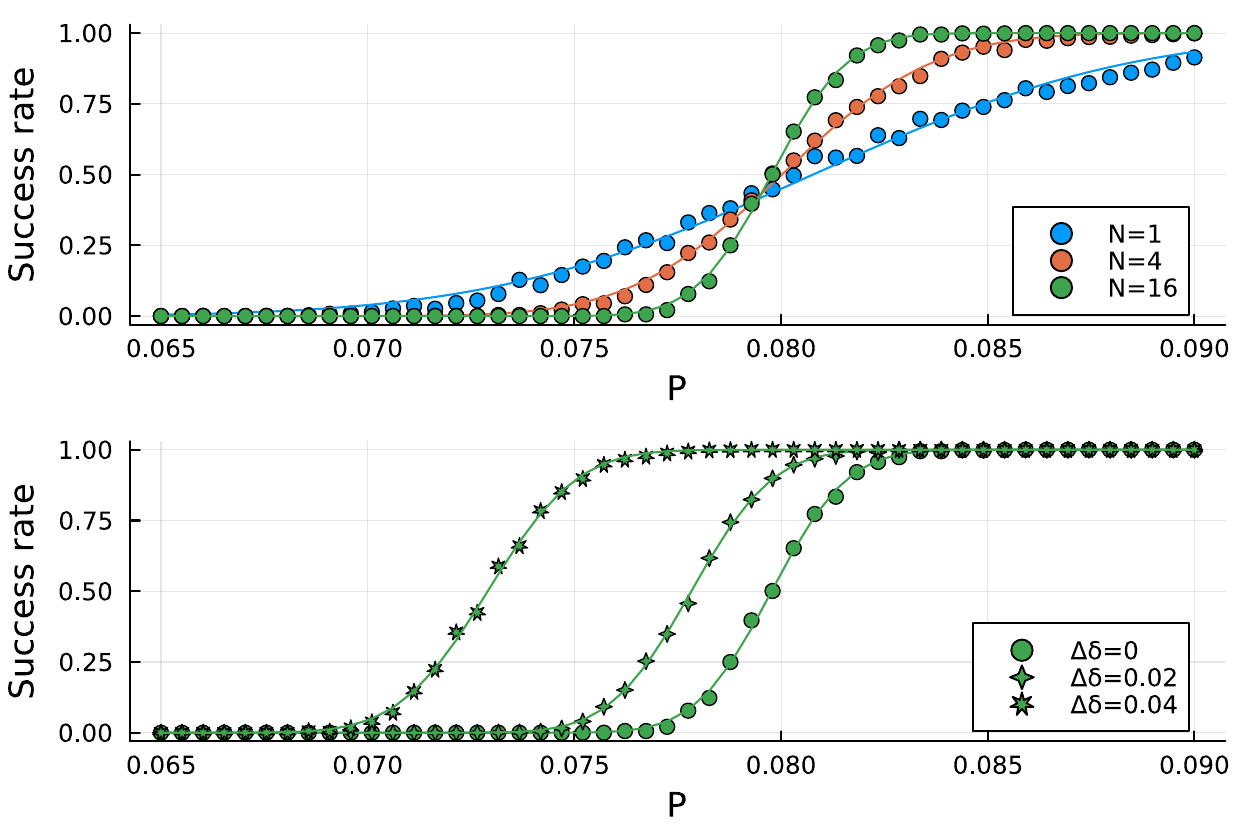}
	\setlength{\unitlength}{1cm}
	\caption{Numerical observation of the efficiency of a perturbation in triggering a spike in presence of noise only in the fields. Top: larger homogeneous networks ($\Delta\delta=0$) show a steeper response. Bottom: for a fixed network size (here $N=16$), the width of the distribution of the $\delta_i$ noticeably modifies the excitability threshold but has almost no impact on the steepness. The other parameters are $(\epsilon = 2.10^{-4}, \alpha_H = 4.0, k = 1.7, \delta = 0.95, \Delta\delta=0.08, \gamma = 0.004, A = 1.43, \sigma = 0.001, \alpha_f = 0.9, \sigma_w=0)$.
	\label{fig:respcurve_num}}
\end{figure}

On \ffig{fig:respcurve_num}, we measure the probability to trigger a spike depending on the perturbations strength P. On the top panel, a single node is compared to two networks with $N=4$ and $N=16$. For each network and for each value of $P$, we perform 1000 independent realizations of applying a single perturbation to the network and count the proportion of successful spike generation. For all network sizes, we find that the dependence of the probability is well fit by the sigmoid function introduced in \ref{eq:sigmoid}. However, contrary to the experiment (\ffig{fig:respcurve}), we observe that the response curve becomes markedly steeper when the network grows, with $\lambda_{N=1}=115, \lambda_{N=4}=245, \lambda_{N=16}=490$. This observation is realized with populations of $N$ identical lasers but we have checked that this phenomenon persists even in the case of diversity in the population. This is shown on the bottom panel of \ffig{fig:respcurve_num}, where we compare the response for different populations of identical size $N=16$ but different distributions of $\delta_i$ characterized by $\Delta\delta=(0, 0.02, 0.04)$ such that $\delta_i=\delta + \Delta\delta (\frac{i-1}{N-1} - \frac{1}{2})$. We see that a population with a broader distribution has a lower excitability threshold (the whole curve is shifted towards smaller values of $P$) but the slope is only minimally affected ($\lambda=490,457,416$ for $\Delta\delta=0,0.02,0.04$ respectively. We note that all moments of the distribution of the $\delta_i$ (and not only mean or variance) actually have an impact on the excitability threshold $P_0$ but the impact on the steepness $\lambda$ is always very small. From these observations, we conclude that a larger network (even a heterogeneous one) is less sensitive than a smaller network to the presence of uncorrelated spontaneous emission noise. However, this does not match the experimental observation (\ffig{fig:respcurve}).

\begin{figure}[h]
	\setlength{\unitlength}{1cm}
	\begin{picture}(8, 6)(0, 0)
		\put(0,0){\includegraphics[width=0.5\textwidth]{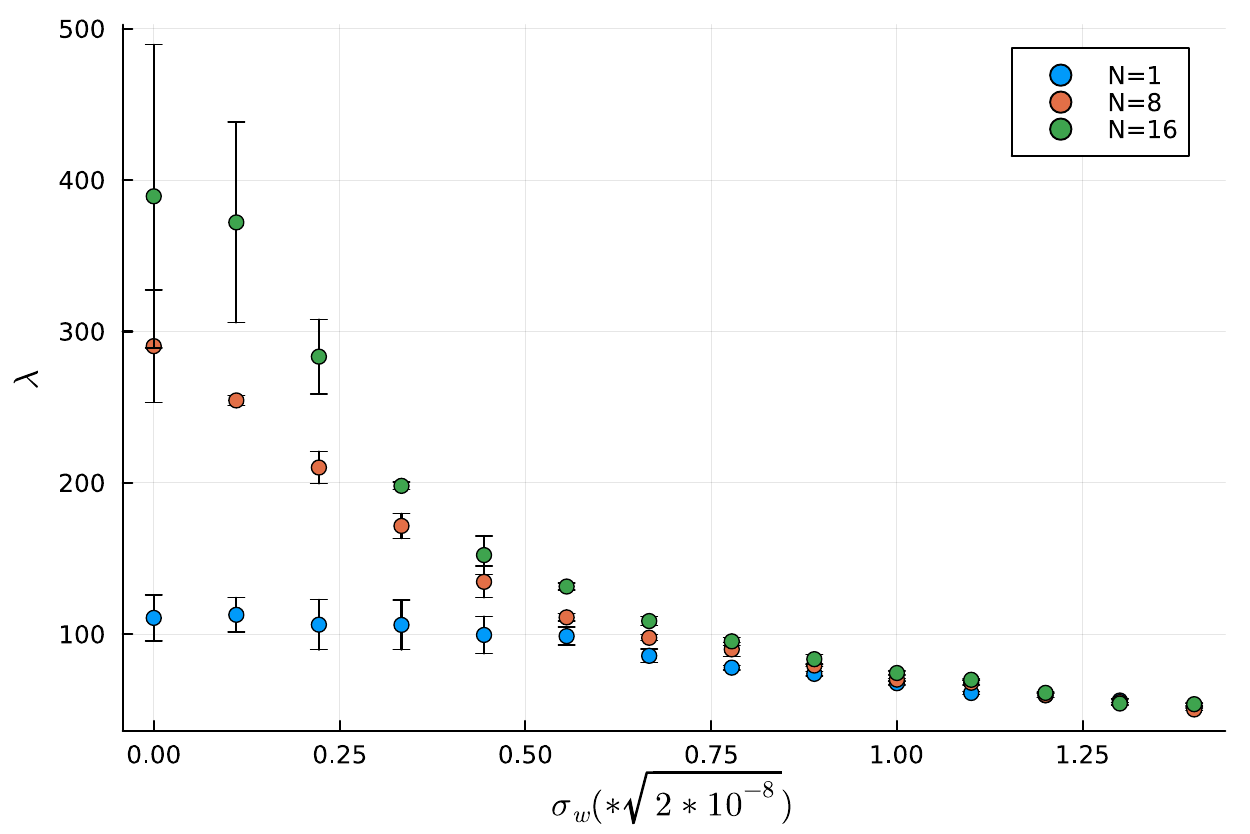}}
		\put(3,2.5){\includegraphics[width=0.23\textwidth]{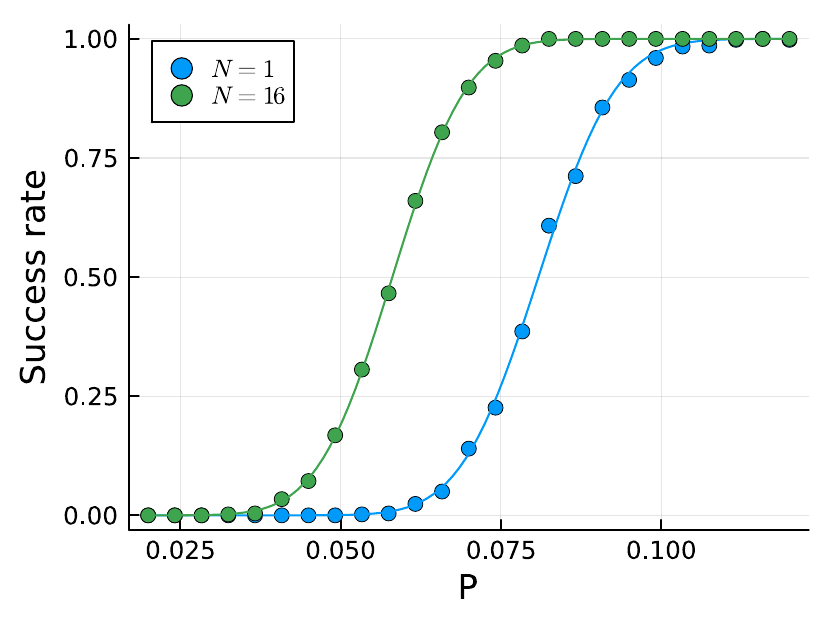}}
        \end{picture}
	\caption{Influence of global noise $\sigma_w$ on the response of networks to perturbations. When $\sigma_w$ is small, large networks have a much steeper response curve. On the contrary, when $\sigma_w$ is large enough, networks of different sizes $N=1,8,16$ display identically steep response curves as illustrated in the inset (to be compared to the experimental observation \ffig{fig:respcurve}). The parameters are $(\epsilon = 2.10^{-4}, \alpha_H = 4.0, k = 1.7, \delta = 0.95, \Delta\delta=0.08, \gamma = 0.004, A = 1.43, \sigma = 0.001, \alpha_f = 0.9)$. For the inset, $\sigma_w=0.9*\sqrt{2*10^{-8}}$. Similar results are obtained when $\Delta\delta=0$.
	\label{fig:sigmaw}}
\end{figure}

In fact, we have checked numerically that this reduced sensitivity of larger networks diminishes when the noise sources $B_i$ are correlated (and even disappears in the case of fully correlated noises). Thus, correlated noise sources could provide an explanation for the experimental observation of \ffig{fig:respcurve}. This is however physically not plausible since the Brownian processes $B_i$ model spontaneous emission. As we show on \ffig{fig:sigmaw}, noise in the adaptive coupling variable $w$ provides a source of correlated noise to the whole network and allows for the reproduction of the experimental observations. Here we measure the slope of the response curve depending on the amount of noise $\sigma_w$ in the coupling variable $w$. This measurement is realized by performing 500 independent realizations of applying a perturbation and counting the proportion of successful spike generation for 25 different values of $P$. This procedure is repeated as a function of the global noise $\sigma_w$ and each resulting curve is then fit with the usual error function (\ref{eq:sigmoid}). We report the measurements in \ffig{fig:sigmaw} for different network sizes $N=1,8,16$, all other parameters being equal. As observed earlier, when $\sigma_w=0$ the response curve is steeper (larger value of $\lambda$) when the network is larger. When $\sigma_w$ increases however the differences diminish and for $\sigma_w\gtrapprox0.75$ the value of $\lambda$ becomes essentially independent of the network size $N$. Thus, the presence of a non-zero global noise $\sigma_w$ is key to reproducing the experimental observation of \ffig{fig:respcurve}. This is shown on the inset of \ffig{fig:sigmaw}, where we show two response curves with identical slopes for a single node with feedback and an $N=16$ network.

\begin{figure}[h]
	\includegraphics[width=0.5\textwidth]{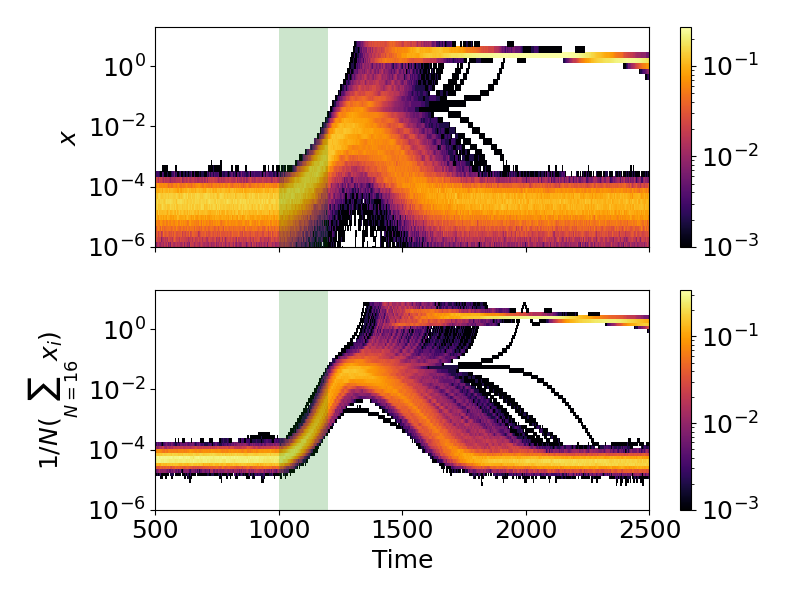}
	\setlength{\unitlength}{1cm}
	\caption{Numerical observation of near threshold distribution of the trajectories of a single node (top) and a mean intensity coupled network (bottom) in response to perturbation with an approximately 0.5 efficiency. The parameters are $(\epsilon = 2.10^{-4}, \alpha_H = 4.0, k = 1.7, \delta = 0.95, \Delta\delta=0.08, \gamma = 0.004, A = 1.43, \sigma = 0.001, \alpha_f = 0.9, \sigma_w=0.7*\sqrt{2*10^{-8}})$.
	\label{fig:departure_num}}
\end{figure}

Including both uncorrelated and global sources of noise, we can reproduce the experimentally observed time traces when triggering an excitable spike either for a single node with feedback and for a mean intensity coupled network with N=16 as shown on \ffig{fig:departure_num}. We have checked that the very long transients where the system remains close to the separatrix (both for $N=1$ and $N=16$) are only mildly affected by the global noise term $\sigma_w$ and that these transients persist for all accessible values of $\sigma_w$.

\section{Discussion}
\label{sec: discussion}
As we have seen above, different sources of noise must be included in the model in order to account for the experimental observations. First, uncorrelated noise sources in the $x_i$ node variable are necessary for the network to ever leave the initial state $x_i=0$, which, even if unstable, is always a fixed point of the system. The network coupled through the mean intensity is globally much less sensitive to these sources of noise than single nodes with feedback. This can be understood by noting that the synchronization manifold is always stable anywhere along the slow manifold. Since the $2N-1$ dimensional stable manifold of the saddle point plays the role of a separatrix for the spiking process at the network level, noise affects the emission of a spike only along the symmetric direction $x_i=\langle x \rangle$, which leads to an effective noise variance of $\sigma^2/N$. Thus, reproducing the experimental observation of a very smooth response curve even for large networks requires the introduction of a correlated noise source. The only physically plausible mechanism for this is electronic noise $\sigma_w$ in the $w$ variable, which adaptively couples all lasers. Since $w$ parametrizes the separatrix, the introduction of $\sigma_w$ effectively introduces a stochastic shift of the latter in the course of time, explaining the smooth response curve of large networks. It is remarkable that driving the coupling of this adaptive network with a random process has an equivalent effect to introducing correlations in the stochastic terms of each node, effectively enhancing the noise component along the synchronization manifold which drives the emission of a spike at the network level.

\section{Conclusion}
\label{sec: summary}
Summarizing, we have analyzed experimentally and numerically the dynamics of an adaptive network of spiking nodes, focusing on the role of noise sources in sum and mean-intensity coupling configurations. The individual nodes are not excitable, but, in the former case, the network can become excitable when it is large enough. In the latter case, we have discussed how coupling can restore excitability at the network level, in contrast to the case of a an ensemble of excitable nodes with feedback. We have examined how network heterogeneity shifts the network response curve without altering its slope, leading us to consider the impact of node and coupling noise terms to fully reproduced the experimental observations.

The overall dynamics is explained by the $2N-1$-dimensional stable manifold of a symmetric $x_i=\langle x \rangle, y_i=\langle y \rangle$ saddle point separating the phase space for the slow manifold of the network, which we could show analytically in the homogeneous case. A further understanding of the dynamics of this adaptive network should be completed by studying the Canard orbits \cite{balzer2024canard,d2021canard} that form the intricate branches of the slow manifold, and their interplay with different noise sources.

\section{acknowledgments}
O.D. thanks V. Kotolup, M. Astner, and L. Pompigna (Maastricht University) for many useful discussions in the framework of their Bachelor/Master thesis research.

\section{References}
\bibliography{bibnet}%
\appendix
\section{Linear stability analysis network}
We recall that $f(0)=0$.
The fast flow of a population of $N$ identical lasers, with $w=0$, is modelled by
\begin{eqnarray}
\dot{x}_i=x_i(y_i-1)\nonumber\\
\dot{y_i}=\gamma(\delta-y+kAf(\langle x \rangle)-xy)
\end{eqnarray}
The system has an OFF equilibrium $X_0^{(N)}$, with $x_i=0, y_i=\delta$ and two symmetric ON equilibria $X_1^{(N)}$ and $X_{sl}^{(N)}$, with $y_i=1$. Besides these symmetrical equilibria, the system has a number of asymmetric equilibrium points, where some lasers are ON and others are OFF. These are given by $(x_i,y_i)=(\langle x \rangle_{ON},1)$, with $\langle x\rangle_{ON} = \langle\delta\rangle_{ON}-1+k(w+Af(\frac{N^+}{N}\langle x \rangle)$ for the ON lasers. Here, $N^+$ denotes the number of lasers that are ON, and the suffix ON denotes an average over those lasers, $\langle x \rangle = \frac{N^+}{N}\langle x \rangle_{ON}$. Note that, depending on $N^+$, there can be one or two roots for $\langle x \rangle_{ON}$. For the OFF lasers, we find $(x_i,y_i)=(0,\delta_i+kAf(\langle x \rangle)$. 

The stability of the equilibria is calculated as follows. The Jacobian of the network can be written as
\begin{equation}
J=\left(\begin{array}{cccc}
   J_{S,1}  &  J_C & J_C & \cdots\\
   J_C  & J_{S,2} & J_C & \ddots \\
   \vdots & & \ddots & \\
\end{array}\right)
\end{equation}
with 
\begin{equation}
J_{S,i}=\left(\begin{array}{cc} y_i-1 & x_i \\
\gamma(\frac{kA}{N}f'(\langle x \rangle)-y_i) & -\gamma(1+x_i)\end{array}\right)
\end{equation}
and 
\begin{equation}
J_C=\left(\begin{array}{cc} 0 & 0 \\
\gamma\frac{kA}{N}f'(\langle x \rangle) & 0\end{array}\right)
\end{equation}
For any laser that is OFF, we find eigenvalues $\lambda_{1,i} =y_i-1$ and $\lambda_{2,i}=-\gamma$. Since $y_i=\delta-1+kw$ for the OFF fixed point, this equilibrium is stable for $\delta<1$ and $w\approx 0$.

For the asymmetric fixed points, this results in $\lambda_{1,i}=\langle x \rangle_{ON} > 0$; all these asymmetric equilibria have hence at least one unstable direction per OFF laser. 

For the symmetric ON fixed points, we evaluate the Jacobian along the symmetric direction 
$$\left(\begin{array}{c}x_{sym}\\y_{sym}\end{array}\right)=\left(\begin{array}{c}x\\y\end{array}\right)\bigotimes\left(\begin{array}{c} 1\\1\\ \vdots \end{array}\right)\,.$$ This results in a Jacobian $J_{sym}=J_S+(N-1)J_C=J_1$, the Jacobian of a single laser. In the symmetric direction, the stability is the same as for a single laser.

In the transverse directions, 
$$\left(\begin{array}{c}x_{T,k}\\y_{T,k}\end{array}\right)=\left(\begin{array}{c}x\\y\end{array}\right)\bigotimes\left(\begin{array}{c} 1\\e^{\frac{2 k\pi i}{N}}\\ \vdots \end{array}\right)$$, with $k=1,\hdots, N-1$, we find a transverse Jacobian $J_T=J_S-J_C=\left(\begin{array}{cc} y-1 & x \\
-\gamma y & -\gamma(1+x)\end{array}\right)$. For $y=1$ (the ON fixed points), the transverse directions are always stable. 

Hence, we have one stable ON fixed point $X^{(N)}_1$. The other fixed point $X^{(N)}_{sl}$ is a saddle, with one unstable direction along the symmetric direction, and $2N-1$ stable directions. All the transverse directions are part of the stable manifold in the neighborhood of the saddle, and hence the separatrix.

This picture remains qualitatively the same for slightly different $\delta_i$\cite{kato2013perturbation}.
\end{document}